\documentclass[12pt,fleqn]{article}
\usepackage{graphicx}

\usepackage{latexsym}

\usepackage{amsmath}
\usepackage{amsthm}
\usepackage{amssymb}
\usepackage{amsfonts}

\begin{document}

\newcommand{\rf}[1]{(\ref{#1})}
\newcommand{\rff}[2]{(\ref{#1}\ref{#2})}

\newcommand{\ba}{\begin{array}}
\newcommand{\ea}{\end{array}}

\newcommand{\be}{\begin{equation}}
\newcommand{\ee}{\end{equation}}

\newcommand{\const}{{\rm const}}
\newcommand{\ep}{\varepsilon}
\newcommand{\Cl}{{\cal C}}
\newcommand{\rr}{{\vec r}}
\newcommand{\ph}{\varphi}
\newcommand{\R}{{\mathbb R}}  
\newcommand{\C}{{\mathbb C}}  
\newcommand{\T}{{\mathbb T}}  
\newcommand{\Z}{{\mathbb Z}}  
\newcommand{\Q}{{\mathbf Q}}
\newcommand{\PP}{{\mathbf P}}

\newcommand{\e}{{\bf e}}

\newcommand{\m}{\left( \ba{c}}
\newcommand{\ema}{\ea \right)}
\newcommand{\mm}{\left( \ba{cc}}
\newcommand{\miv}{\left( \ba{cccc}}

\newcommand{\scal}[2]{\mbox{$\langle #1 \! \mid #2 \rangle $}}
\newcommand{\ods}{\par \vspace{0.5cm} \par}
\newcommand{\dis}{\displaystyle }
\newcommand{\mc}{\multicolumn}

\newtheorem{prop}{Proposition}
\newtheorem{Th}[prop]{Theorem}
\newtheorem{lem}[prop]{Lemma}
\newtheorem{rem}{Remark}
\newtheorem{cor}[prop]{Corollary}
\newtheorem{Def}{Definition}
\newtheorem{open}{Open problem}
\newtheorem{ex}{Example}
\newtheorem{exer}{Exercise}

\newenvironment{Proof}{\par \vspace{2ex} \par
\noindent \small {\it Proof:}}{\hfill $\Box$ 
\vspace{2ex} \par }

\title{\bf 
Comment on `conservative discretizations of the Kepler motion' }
\author{
 {\bf Jan L.\ Cie\'sli\'nski}\thanks{\footnotesize
 e-mail: \tt janek\,@\,alpha.uwb.edu.pl}
\\ {\footnotesize Uniwersytet w Bia{\l}ymstoku,
Wydzia{\l} Fizyki}
\\ {\footnotesize ul.\ Lipowa 41, 15-424
Bia{\l}ystok, Poland}
}

\date{}

\maketitle

\begin{abstract}
We show that the exact integrator for the classical Kepler motion, recently found by Kozlov ({\it J.\ Phys.\ A: Math.\ Theor.\ } {\bf 40} (2007) 4529-4539), can be derived in a simple natural way (using well known exact discretization of the harmonic oscillator). We also turn attention on important earlier references, where the exact discretization of the 4-dimensional isotropic  harmonic oscillator has been applied to the perturbed Kepler problem. 
\end{abstract}

\ods

{\it PACS Numbers:} 45.10.-b; 45.50.Pk; 02.70.Bf; 02.60.Cb; 02.30Hq; 02.30Ik

{\it MSC 2000:} 65P10; 65L12; 34K28

{\it Key words and phrases:} geometric numerical integration, Kepler problem, Kustaanheimo-Stiefel transformation, harmonic oscillator, integrals of motion, exact numerical integrators

\pagebreak

In recent years there were proposed several conservative discretizations of the classical Kepler problem \cite{QD,MN1,MN2,Koz,Ci-Kep}. These numerical integrators preserve all integrals of motion and trajectories but only Kozlov's schemes are of order higher than 2. Kozlov found also the exact integrator by guessing its proper form and summing up some infinite series \cite{Koz}.

In this comment we show that Kozlov's exact integrator can be derived in a simple elementary way. 
Conservative discretizations of the 3-dimensional Kepler motion obtained in \cite{QD,MN2,Koz} consist in applying the midpoint rule (or the discrete gradient method, compare \cite{LaG}) to the isotropic 4-dimensional harmonic oscillator equations:
\be  \label{KS-osc}
  \frac{d \Q}{d s} = \frac{1}{4} \PP \ , \quad \frac{d \PP}{d s} = 2 E \Q \ , \qquad ( \Q, \PP \in \R^4 )
\ee
 where $E = \const$ is the energy integral of the considered Kepler motion. Then, the Authors of \cite{MN2,Koz} use the Kustaanheimo-Stiefel  (KS) transformation. This classical transformation is given by (\cite{KS}, see also \cite{Koz}):  
\be \label{Hopf}
\mathbf q = \m 
Q_1^2 - Q_2^2 - Q_3^2 + Q_4^2 \\
  2 Q_1 Q_2 - 2 Q_3 Q_4  \\
 2 Q_1 Q_3 + 2 Q_2 Q_4 
  \ema , 
\ee
\be  \label{KS-P}
\mathbf p = \frac{1}{ 2 |\Q|^2 } \m 
P_1 Q_1 - P_2 Q_2 - P_3 Q_3 + P_4 Q_4 \\
P_1 Q_2 + P_2 Q_1 - P_3 Q_4 - P_4 Q_3 \\ 
P_1 Q_3 + P_2 Q_4 + P_3 Q_1 + P_4 Q_2 
\ema  ,
\ee
where $\Q, \PP$ are subject to the constraint
\be \label{constr}
P_1 Q_4 - P_2 Q_3 + P_3 Q_2 - P_4 Q_1 = 0 \ .
\ee
The KS transformation, together with the Levi-Civita time transformation
\be  \label{LCDS} 
\frac{d t}{d s} = | {\mathbf q}| 
\ee
maps the 4-dimensional harmonic oscillator \rf{KS-osc} into the 3-dimensional Kepler problem equations:  
\be  \label{Kep-qp}
 \frac{d {\mathbf q} }{d t} = {\mathbf p} \ , \quad 
\frac{d {\mathbf p} }{d t} = - \frac{k {\mathbf q}}{|{\mathbf q}|^3}  \ , \qquad ( {\mathbf q, \mathbf p} \in \R^3 ) \ ,
\ee
where $k = \const$. Using \rf{Hopf}, \rf{KS-P} and \rf{constr} we can verify useful identities  
\be
  |\mathbf q|^2 = |\Q|^4 \ , \quad |\PP|^2 = 4 |\mathbf p|^2  |\Q|^2 \ ,
\ee
which imply the equivalence of the energy conservation laws:
\be  \label{conlaw}
\frac{1}{2} {\mathbf p}^2 - \frac{k}{|\mathbf q|} = E  \qquad 
\Longleftrightarrow \qquad  \frac{1}{8} |\mathbf P|^2 - E |\mathbf Q|^2 = k \ .
\ee

The phenomenon of interchanching coupling constants with integrals of motion (like $k \leftrightarrow E$) is quite well known in the theory of integrable systems, see \cite{HGDR} (compare also  \cite{Se-Bl}, where more general results can be found). 

 \ods

In order to derive Kozlov's numerical results in a simple straightforward way 
it is sufficient to notice that the KS transformation  (used by Kozlov), reduces the Kepler motion to linear ordinary differential equations with constant coefficients (namely, to the harmonic oscillator) and for all such equations there exist explicit exact numerical integrators (\cite{Po,Ag}, see also \cite{Ci-oscyl}). By the exact discretization of an ordinary differential equation $\dot x = f (x)$, where $x (t) \in \R^N$, we mean the one-step numerical scheme of the form $X_{n+1} = \Phi_h (X_n)$, such that $X_n = x (t_n)$, compare \cite{Po,Ag}.

The system \rf{KS-osc}, equivalent to the 4-dimensional harmonic oscillator equation, admits the exact discretization (see, for instance, \cite{Ci-oscyl}):
\be  \ba{l}  \dis  \label{exact-mid}
\frac{\Q_{j+1} - \Q_j }{ \delta (h_j) } = \frac{1}{4} \  \frac{\PP_{j+1} + \PP_j}{2}  \ , \\[3ex]\dis
\frac{\PP_{j+1} - \PP_j }{ \delta (h_j) } = 2 E \ \frac{\Q_{j+1} + \Q_j}{2} \ ,
\ea \ee
where $h_j := s_{j+1} - s_j$ is the (variable) $s$-step, $\Q_j, \PP_j$ denote $j$th iteration of the numerical scheme (not to be confused with coordinates $Q_j, P_j$), and  
\be
 \delta (h_j) = \frac{2}{\omega}  \tan \frac{\omega h_j}{2} \ , \qquad \omega^2 = - \frac{1}{2} E \ .
\ee
In the case of the constant step $ h_j = h$ and $E < 0$, we recognize here 
the exact integrator found by Kozlov (see formulae (4.11) and (4.14) from \cite{Koz}, taking into account that $\delta (h) = h a (h) = h b(h)$ and $E = - A$). The hyperbolic and parabolic cases (formulae (4.16) and (4.18) from  \cite{Koz}) follow immediately when we take imaginary $\omega$ (i.e., $E >0$) or $\omega = 0$, respectively. 
The exact numerical scheme \rf{exact-mid} preserves the energy integral, i.e., 
\be \label{energy}
\frac{1}{8} |\mathbf P_j|^2 - E |\mathbf Q_j|^2 = k \ .
\ee
Note that the system \rf{exact-mid} can be rewritten in the explicit form: 
\be \ba{l} \dis  \label{exact-explicit}
\Q_{j+1} = \cos\omega h_j \ \Q_j + \frac{\sin\omega h_j}{4 \omega} \ \PP_j \ , \\[2ex] \dis 
\PP_{j+1} = - 4 \omega \sin\omega h_j \ \Q_j + \cos\omega h_j \ \PP_j 
\ . \ea \ee
This system is a direct consequence of evaluating the exact solution of \rf{KS-osc} at $s = s_j$ and $s = s_j + h_j$, compare \cite{Ci-oscyl}. 

\ods
The  equation \rf{LCDS} can be solved exactly in different (but more or less equivalent) ways, compare \cite{Koz,Mi,Br}. Here we propose one more approach, reducing this problem to linear ordinary differential equations with constant coefficients.  
If $\Q, \PP$ satisfy \rf{KS-osc} and $t$ satisfies \rf{LCDS}, then we easily check that
\be   \label{w-Omega}
\frac{d \mathbf w}{d s} = \Omega \mathbf w , \quad \mathbf w = \m |\Q|^2  \\ |\PP|^2 \\ \Q \cdot \PP \\ t \ema   , \quad    
\Omega = \left( \ba{cccc} 
0 & 0 & \frac{1}{2} & 0 \\ 0 & 0 & 4 E & 0 \\ 2 E & \frac{1}{4} & 0 & 0  \\ 1 & 0 & 0 & 0 \ea \right)  . 
\ee
In such case we can proceed in a standard way. The 
general solution is given by $\mathbf w (s) = \exp (s \Omega) \mathbf w (0)$. Therefore, the exact discretization, $\mathbf w_n = \mathbf w ( h n)$, satisfies
\be  \label{wom} 
 \mathbf w_{n+1} = \exp ( h \Omega) \mathbf w_n \ , 
\ee
and the problem reduces to the well known, purely algebraic procedure of computing $e^{\Omega h}$. In our particular case we observe that 
$\Omega^4 = 2 E \Omega^2$ which simplifies computations. 
The last row in the equation \rf{wom} reads
\be  \label{tj+1}
t_{j+1} = t_j + \frac{\sin 2 h \omega}{4 \omega} \left( |\Q_j|^2  - \frac{|\PP_j|^2 }{16 \omega^2} \right) + \frac{h}{2}  \left( |\Q_j|^2  + \frac{|\PP_j|^2 }{16 \omega^2} \right) + 
 \frac{\Q_j \cdot \PP_j \sin^2 h\omega}{4 \omega^2} .
\ee
One can check by direct computation that the discretization \rf{tj+1}, although have a simpler form,  is  identical with the formulae (4.11), (4.15) of \cite{Koz}. 
Finally, eliminating $|\PP_j|^2$ by virtue of \rf{energy}, we get 
\be  \label{time}
t_{j+1} = t_j + \frac{h k}{4 \omega^2} \left( 1 - \frac{\sin 2 h \omega}{2 h \omega} \right) + \frac{\sin 2 h \omega}{2 \omega} |\Q_j|^2  + \frac{\Q_j \cdot \PP_j \sin^2 h\omega}{4 \omega^2} . 
\ee
Another approach (see \cite{Mi}) consists in computing the integral $ \int |\Q (s)|^2 ds$, where $\Q$  is the exact solution of \rf{KS-osc}.  The formula (86) from \cite{Mi} is identical to \rf{time} (although notation is quite different).

In celestial mechanics the exact discretization of the Kepler motion {\it via} the KS transformation appeared as a quite natural step \cite{Mi,Br,SB}, 
although the conservative properties of the exact integrator were not discussed explicitly in these papers. 
A long time ago 
Stiefel and Bettis, working in the framwork of the Gautschi approach \cite{Gau}, applied  
the exact discretization of the harmonic oscillator to the perturbed Kepler motion \cite{SB,Be2}. 

More recently, Mikkola \cite{Mi} and Breiter \cite{Br} proposed new integrators for the perturbed Kepler problem, using  the exact solution of the 4-dimensional isotropic harmonic oscillator equation and the exact discretization \rf{time} of the time (known to Stumpff even before the KS transform was invented, compare \cite{Mi}). 
In particular, the numerical scheme \rf{exact-explicit} can be found in 
\cite{Mi}, p.162, and in \cite{Br}, p.234.  Breiter follows \cite{DEF} using an additional constant in the definition of the KS transformation (in fact scaling both $\mathbf q$ and $\mathbf p$). The freedom of choosing this parameter can be used to fix the value of $\omega$ (e.g., $\omega = 1$) which may have numerical advantages \cite{Br}. 

These important results of celestial mechanics are not mentioned in \cite{Koz}   and, in general,  they seem to be rather unknown in the field of geometric numerical integration \cite{HLW}. 
It is worthwhile to mention that the exact discretization of the harmonic oscillator equation has been recently used to construct new geometric integrators of high accuracy (``locally exact discrete gradient schemes'') \cite{CR-grad}. We plan to apply such scheme to the perturbed Kepler problem using the Kustaanheimo-Stiefel map.



\begin{thebibliography}{99}

\newcommand{\vbib}{ \par \vspace{-2ex} \par \bibitem}

\footnotesize

\vbib{QD}
G.R.W.Quispel, C.Dyt: 
``Solving ODE's numerically while preserving symmetries, hamiltonian structure, phase space volume, or first integrals'', 
{\it Proc.\ 15th IMACS World Congress}, vol.\ II, ed.\ by A.Sydow, pp.\ 601-607; Wissenschaft \& Technik Verlag, Berlin 1997.

\vbib{MN1}
Y.Minesaki, Y.Nakamura:
``A new discretization of the Kepler motion which conserves the Runge-Lenz vector'', 
{\it Phys.\ Lett.\ A} {\bf 306} (2002) 127-133.



\vbib{MN2}
Y.Minesaki, Y.Nakamura: 
``A new conservative numerical integration algorithm for the three-dimensional Kepler motion based on the Kustaanheimo-Stiefel regularization theory'', 
{\it Phys.\ Lett.\ A} {\bf 324} (2004) 282-292. 


\vbib{Koz}
R.Kozlov: 
``Conservative discretizations of the Kepler motions'', 
{\it J.\ Phys. A: Math.\ Theor.} {\bf 40} (2007) 4529-4539.



\vbib{Ci-Kep}
J.L.Cie\'sli\'nski:
``An orbit-preserving discretization of the classical Kepler problem'',
{\it Phys.\ Lett.\ A} {\bf 370} (2007) 8-12.



\vbib{LaG} 
R.A.LaBudde, D.Greenspan: 
``Discrete mechanics -- a general treatment'', 
{\it J.\ Comput.\ Phys.\ } {\bf 15} (1974) 134-167.

\vbib{KS}
P.Kustaanheimo, E.Stiefel: 
``Perturbation theory of Kepler motion based on spinor regularization'', 
{\it J.\ reine angew. Math.\ } {\bf 218} (1965) 204-219. 

\vbib{HGDR} 
J.Hietarinta, B.Grammaticos, B.Dorizzi, A.Ramani: 
``Coupling-constant metamorphosis and duality between integrable Hamiltonian systems'', 
{\it Phys.\ Rev.\ Lett.\ } {\bf 53} (1984) 1707-1710.


\vbib{Se-Bl} 
A.Sergyeyev, M.B{\l}aszak: 
``Generalized St\"ackel transform and reciprocal transformation for finite-dimensional integrable systems'', 
{\it J.\ Phys. A: Math.\ Theor.} {\bf 41} (2008) 105205. 


\vbib{Po}
R.B.Potts: 
``Differential and difference equations'', 
{\it Am.\ Math.\ Monthly} {\bf 89} (1982) 402-407. 



\vbib{Ag}
R.P.Agarwal:
{\it Difference equations and inequalities} (Chapter 3),
Marcel Dekker, New York 2000.


\vbib{Ci-oscyl}
J.L.Cie\'sli\'nski:
``On the exact discretization of the classical harmonic oscillator equation'', 
{\it preprint arXiv}: 0911.3672v1 [math-ph] (2009).  


\vbib{Mi}
S.Mikkola: 
``Practical symplectic methods with time transformations for the few-body problem'', 
{\it Celestial Mech. Dyn.\ Astron.\ } {\bf 67} (1997) 145-165.


\vbib{Br}
S.Breiter:
``Explicit symplectic integrator for highly eccentric orbits'', 
{\it Celestial Mech. Dyn.\ Astron.\ } {\bf 71} (1999) 229-241.


\vbib{SB}
E.Stiefel, D.G.Bettis:
``Stabilization of Cowell's method'', 
{\it Numer.\ Math.\ } {\bf 13} (1969) 154-175.


\vbib{Be2}
D.G.Bettis:
``Stabilization of finite difference methods of numerical integration'', 
{\it Celestial Mech.\ } {\bf 2} (1970) 282-295. 


\vbib{Gau}
W.Gautschi: 
``Numerical integration of ordinary differential equations based on trigonometric polynomials'', 
{\it Numer.\ Math.} {\bf 3} (1961) 381-397.

\vbib{DEF}
A.Deprit, A.Elipe, S.Ferrer: 
``Linearization: Laplace vs.\ Stiefel'', 
{\it Celestial Mech. Dyn.\ Astron.\ } {\bf 58} (1994) 151-201.



\vbib{HLW}
E.Hairer, C.Lubich, G.Wanner:
{\it Geometric numerical integration: structure-preser\-ving algorithms for ordinary differential equations},
Second Edition, Springer, Berlin 2006.



\vbib{CR-grad}
J.L.Cie\'sli\'nski, B.Ratkiewicz:
``How to improve the accuracy of the discrete gradient method in the one-dimensional case'',  
{\it preprint arXiv}: 0901.1906v1 [cs.NA] (2009).  
\par \vspace{-1ex} \par 
J.L.Cie\'sli\'nski, B.Ratkiewicz:
``Improving the accuracy of the discrete gradient method in the one-dimensional case'',  {\it Phys.\ Rev.\ E}, in press.











\end{thebibliography}
\end{document}